\documentclass[pra,aps,onecolumn,twoside,showpacs,superscriptaddress]{revtex4}
\usepackage{amsfonts}
\usepackage{amssymb}
\usepackage{bm, bbm, dsfont}
\usepackage[dvips]{graphicx}
\usepackage[T1]{fontenc}
\usepackage[latin2]{inputenc}
\usepackage{amsmath}
\usepackage{latexsym}
\usepackage{graphics}
\usepackage{subfig}

\usepackage{color}

\def\dt{{\rm d}\,}

\newcommand{\ket}[1]{| #1 \rangle}
\newcommand{\bra}[1]{\langle #1 |}

\def\duzomniejsze{<\kern-.7mm<}
\def\duzowieksze{>\kern-.7mm>}

\def\textbf#1{{\bf #1}}
\def\be{\begin{equation}}
\def\ee{\end{equation}}
\def\ben{\begin{eqnarray}}
\def\een{\end{eqnarray}}
 \def\beqa{\begin{eqnarray}}
\def\eeqa{\end{eqnarray}}
\def\eea{\end{array}}
\def\bea{\begin{array}}
\newcommand{\bei}{\begin{itemize}}
\newcommand{\eei}{\end{itemize}}
\newcommand{\bee}{\begin{enumerate}}
\newcommand{\eee}{\end{enumerate}}

\def\1{\openone}

\def\>{\rangle}
\def\<{\langle}

\def\dt#1{{{\kern -.0mm\rm d}}#1\,}
\def\squareforqed{\hbox{\rlap{$\sqcap$}$\sqcup$}}
\def\qed{\ifmmode\squareforqed\else{\unskip\nobreak\hfil
\penalty50\hskip1em\null\nobreak\hfil\squareforqed
\parfillskip=0pt\finalhyphendemerits=0\endgraf}\fi}

\newtheorem{lemma}{Lemma}
\newtheorem{theorem}[lemma]{Theorem}
\newtheorem{main result}[lemma]{Main result}
\newtheorem{proposition}[lemma]{Proposition}
\newtheorem{definition}{Definition}

\newtheorem{fact}[lemma]{Fact}

\def\bep{\begin{proposition}}
\def\eep{\end{proposition}}
\def\bel{\begin{lemma}}
\def\eel{\end{lemma}}

\def\bet{\begin{theorem}}
\def\eet{\end{theorem}}
\def\bed{\begin{definition}}
\def\eed{\end{definition}}
\def\bef{\begin{fact}}
\def\eef{\end{fact}}

\begin{document}
%\language\english

%preprint style
%\documentstyle[preprint,aps]{revtex}

%preprint style - single spaced
%\documentstyle[tighten,aps]{revtex}
% Full title of the paper (Capitalized)
\title{Objectivisation In Simplified Quantum Brownian Motion Models}

% Authors (Add full first names)
\author{J.~Tuziemski}
\affiliation{Faculty of Applied Physics and Mathematics, Gda\'nsk University of Technology, 80-233 Gda\'nsk, Poland}
\affiliation{National Quantum Information Centre in Gda\'nsk, 81-824 Sopot, Poland}
\author{J.~K.~Korbicz}
 \email{jaroslaw.korbicz@ug.edu.pl}
  \affiliation{Institute of Theoretical Physics and Astrophysics, University of Gda\'nsk, 80-952 Gda\'nsk, Poland}
 \affiliation{National Quantum Information Centre in Gda\'nsk, 81-824 Sopot, Poland}

% Abstract (Do not use inserted blank lines, i.e. \\) 
\begin{abstract}Birth of objective properties from subjective quantum world has been one of the key questions in the quantum-to-classical transition.
Basing on recent results in the field, we study it in a quantum mechanical model of a boson-boson interaction---quantum Brownian motion.
Using various simplifications we prove a formation for thermal environments of, so called, spectrum broadcast structures, responsible for perceived objectivity.
In the quantum measurement limit we prove that this structure is always formed, providing the characteristic timescales.
Including self-Hamiltonians of the environment, we show the exponential scaling of the effect with the size of the environment.
Finally, in the full model we numerically study the influence of squeezing in the initial state of the environment, showing broader
regions of formation than for non-squeezed thermal states.
\end{abstract}

%\keyword{decoherence, quantum Brownian motion, broadcasting}

% The fields PACS, MSC, and JEL may be left empty or commented out if not applicable
%\PACS{}
%\MSC{}
%\JEL{}

% If this is an expanded version of a conference paper, please cite it here: enter the full citation of your conference paper, and add $^\dagger$ in the end of the title of this article.
%\conference{}

\maketitle

\section{Introduction}

Quantum measurement inevitably disturbs a state of a measured system, unless very special conditions are met \cite{vN}.
This is in a stark contrast with classical physics we are used to, where a state of the system is believed to exists objectively,
independently of the observation. A pioneering attempt to resolve this situation has been initiated by Zurek and collaborators 
through the quantum Darwinism program (see e.g. \cite{ZurekNature} and the references therein), which is an advance 
form of the theory decoherence (see e.g. \cite{Schlosshauer}). In essence, it relies on two fundamental observations.
First, in usual situations
a quantum system $S$ interacts with multiple environments $E_1,\dots,E_N$ and the information about the system
is learned not directly but through an observation of some portion, which we call $fE$, of those environments.
Hence, studying information content, deposited through decoherence in the portions of environment is of key importance.
Second, what lies at the heart of objectivity is information redundancy: If the same information about the system (or more precisely about 
its state) is present in many fractions of the environment and can be accessed by independent observers without
disturbing the system, it appears as being objective \cite{ZurekNature}. This information redundancy has been studied in several emblematic models of decoherence, 
e.g. in quantum Brownian motion \cite{qbm_Zurek}, in spin-$1/2$ systems \cite{spiny_Zurek}, and 
for a small dielectric sphere illuminated by photons \cite{Zurek_sfera}, suggesting
that the objectivisation of certain properties indeed takes place in those models.

However, this approach is based on a certain scalar criterion for objectivity, probing the scaling of the quantum mutual information
between the system and the observed environment fraction with the size of this fraction \cite{ZurekNature}. Although very suggestive,
the sufficiency of this criterion remains open, as pointed out in \cite{object}, where a deeper structural analysis has been developed,
based on the most fundamental description of a quantum system---its state (see also \cite{generic}, strengthening some aspects
of \cite{object}). In brief, using an operational definition of 
objectivity from \cite{ZurekNature} and, surprisingly, the Bohr non-disturbance condition \cite{Wiseman} it has been 
shown in an abstract, model- and dynamics-independent  setting that the necessary and sufficient condition for appearance
of an objective state of the system is formation of a, so called, {\em spectrum broadcast structure} \cite{object}:
\begin{equation}\label{br2}
\varrho_{S:fE}=\sum_i p_i \ket{\xi_i}\bra{\xi_i}\otimes \varrho^{E_1}_i\otimes\cdots\otimes \varrho^{E_{fN}}_i,\ 
\varrho^{E_k}_i\varrho^{E_k}_{i'\ne i}=0,
\end{equation}
where $fE$ denotes the observed fraction of the environment $E$, $\{\ket{\xi_i}\}$ is a pointer basis, $p_i=\langle \xi_i|\varrho_{0S} |\xi_i\rangle$ 
initial pointer probabilities, and $\varrho^{E_1}_i,\dots,\varrho^{E_{fN}}_i$ some states of the observed environments $E_1,\dots,E_{fN}$
with mutually orthogonal supports. The process of formation of this structure is called {\em state information broadcasting} and is similar to quantum 
state broadcasting \cite{broadcasting}. However, only a part of the initial state $\varrho_{0S}$ is broadcasted---the probabilities  $p_i=\langle \xi_i|\varrho_{0S} |\xi_i\rangle$. The name comes from the fact that all the reduced states of the environments have the same spectrum $\{p_i\}$ as the decohered
state of the system and such structures first appeared in a different context in \cite{my}. As seen from (\ref{br2}), if a spectrum brodcast structure has been formed,
the observers trying to determine the state of the system from the environment will all get the same results $i$ with the same 
probabilities $p_i$, due to the non-overlapping supports of the environmental states,
 and the state of the system will stay unchanged. This leads to the objectivisation of the states $\ket{\xi_i}$.
Formally, state information broadcasting can be described as a quantum channel:
\begin{equation}
\Lambda^{S\to fE}(\varrho_{0S})=\sum_i\langle \xi_i |\varrho_{S0}|\xi_i\rangle \varrho^{E_1}_i\otimes\cdots\otimes \varrho^{E_{fN}}_i.
\end{equation}

Formation of such structures has been recently theoretically confirmed for the mentioned models of decoherence: the illuminated sphere model \cite{sfera}, spin-$1/2$ systems \cite{spiny_my}, and in 
quantum Brownian motion \cite{qbm} (see e.g. \cite{Petruccione,Schlosshauer} for an introduction to the model). 
The last study, although introduced a novel concept of a {\em dynamical objectivisation} with time evolving structure
(\ref{br2}), relied in its final stage on the numerics, due to a complicated nature of appearing expressions.
In the present work, we further analyze quantum Brownian motion model and,  at an expense of various simplifications, we obtain analytical estimates on the 
speed of formation  of the spectrum broadcast structure  as a function of time and
the size of the environment. 
Moreover, in the full model we numerically study the effect of squeezing in the initial state of the 
environment. 

\section{Results}

\subsection{The general mechanism}

We illustrate our method, first developed in \cite{object,sfera}, on a general interaction Hamiltonian of the von Neumann measurement type \cite{vN}:
\begin{equation}\label{vnH}
\hat H =\hat A\otimes \sum_{k=1}^N \hat B_k,
\end{equation}
where $\hat A$, $\hat B_k$ are some observables of the system and the $k$-th environment respectively, assumed for simplicity
to have discrete spectra (for a treatment of observables with continuous spectra see \cite{qbm}) and we have neglected the self-Hamiltonians
of the system and the environments. 
We assume that the strengths of each interaction in (\ref{vnH}) are vanishingly small (one can imagine adding small coupling
constants $g_k$ there; we however do not do it for simplicity's sake).
Diagonalizing $\hat A$, leads to the time evolution of a controlled unitary type, with
the system controlling the environment:
\begin{equation}
\hat U(t)=\sum_a \ket a\bra a\otimes\bigotimes_{k=1}^N e^{-iat\hat B_k}.
\end{equation}
The basic object of our study \cite{object,sfera,qbm} is the post-interaction state $\varrho_{S:fE}(t)$ of the system $S$ and the observed fraction of the environment $fE$ (of the size $fN$, $0<f<1$), 
obtained with the above $\hat U(t)$. The observed fraction is not the whole environment as some losses are necessary for the decoherence (as it happens in realistic situations). We obtain:
\begin{eqnarray}
&&\varrho_{S:fE}(t)=tr_{(1-f)E}\left[\hat U(t)\varrho_{0S}\otimes\bigotimes_{k=1}^N\varrho_{0k}\hat U(t)^\dagger\right]\nonumber\\
&&=\sum_a\langle a|\varrho_{0S}|a\rangle \ket a \bra a\otimes \bigotimes_{k=1}^{fN} e^{-iat\hat B_k}\varrho_{0k}e^{iat\hat B_k}\label{diag}\\
&&+\sum_{a\ne a'} \Gamma_{a,a'}(t) \langle a|\varrho_{0S}|a'\rangle \ket a \bra {a'}\otimes \bigotimes_{k=1}^{fN} e^{-iat\hat B_k}\varrho_{0k}e^{ia't\hat B_k},
\label{off}
\end{eqnarray}
where:
\begin{equation}
\Gamma_{a,a'}(t)=\prod_{k\in (1-f)E} tr\left[e^{-iat\hat B_k}\varrho_{0k}e^{ia't\hat B_k}\right]
\end{equation}
is the usual decoherence factor between the states $\ket a$, $\ket {a'}$ and  $(1-f)E$ denotes the lost (unobserved) fraction of the environment.
We then check for the spectrum broadcast structure (\ref{br2}) formation in two steps.

First, we take care of the vanishing of the off-diagonal part (\ref{off}), which is the well known decoherence process. Vanishing of 
(\ref{off}) in the trace norm is fully controlled (assuming of course coherences between different $a$'s in the system's initial state$\varrho_{0S}$) by
 the modulus $|\Gamma_{a,a'}(t)|$, so it is enough to check if $|\Gamma_{a,a'}(t)|\approx 0$ with time. 
Note that decoherence is a result of the losses and therefore decoherence factor depends only on the unobserved fraction of the environment. 
The basis $\{\ket a\}$ becomes then the pointer basis, which as we pointed out earlier \cite{sfera} is arguably put by hand to some extent.
Developing a framework where this basis appears truly dynamically would be a very interesting task (if at all possible).

Second, we check if the post-interaction states of the environments $\varrho_{ka}(t)=e^{-iat\hat B_k}\varrho_{0k}e^{iat\hat B_k}$
have non-overlapping suports $\varrho_a(t)\varrho_{a'}(t)\approx 0$. This is equivalent to their perfect one-shot distinguishability
through projective measurements on their supports. However, due to the assumed weak interaction (\ref{vnH}) one cannot expect that
for states of the individual environments. In fact, for a vanishingly weak interaction, the states $\varrho_{ka}(t)$ are almost identical for
different $a$'s ans hence carry a vanishingly small amount of information about the system 
(the index $a$ of an eigenstate of the observable $\hat A$ in this case). Thus, we perform an environment coarse-graining
and divide the observed fraction of the environment $fE$ in (\ref{diag}) into $\mathcal M$ fractions (assumed equal for simplicity)
of a size $fN/\mathcal M=mN$. This grouping into macrofractions can be also seen as representing detection thresholds of real-life
detectors. Since at some point we are interested in a thermodynamic-type of a limit $N\to\infty$ it is important
that those fractions scale with $N$ (hence the name 'macrofractions'). Due to no direct interactions between different parts of the environment
(cf. (\ref{vnH})), post-interaction states of each macrofraction are simply products $\varrho_{a}^{mac}(t)=\bigotimes_{k\in mac}\varrho_a(t)$.
We assume that a single macrofraction is a minimal portion of the environment providing the observers with information about the central system and
therefore it is enough to consider the distinguishability at the level of a  single macrofraction.
We  use the generalized overlap \cite{Fuchs}, $B(\varrho_1,\varrho_2)=tr\sqrt{\sqrt{\varrho_1}\varrho_2\sqrt{\varrho_1}}$,
as the most convenient measure due to its scaling with the tensor product:
\begin{equation}
B^{mac}_{a,a'}(t)=B\left(\varrho_{a}^{mac}(t),\varrho_{a'}^{mac}(t)\right)=\prod_{k\in mac}B\left(\varrho_{ka}(t),\varrho_{ka'}(t)\right).
\end{equation}
Vanishing of $B^{mac}_{a,a'}(t)$ is equivalent to $\varrho_{a}^{mac}(t)$, $\varrho_{a'}^{mac}(t)$ being perfectly distinguishable \cite{Fuchs}.

Thus, if one is able to prove that for some initial states $\varrho_{0k}$ (and other possible parameters of the model) both:
\begin{equation}
|\Gamma_{a,a'}(t)|\approx 0,\ B^{mac}_{a,a'}(t)\approx 0,
\end{equation}
then this is equivalent to a formation of the spectrum broadcast structure (\ref{br2}):
\begin{equation}
\varrho_{S:fE}\approx\sum_a p_a \ket a\bra a\otimes \varrho^{mac_1}_a\otimes\cdots\otimes \varrho^{mac_{\mathcal M}}_a,
\ \varrho^{mac_j}_a\varrho^{mac_j}_{a'\ne a}\approx 0,
\end{equation}
and the convergence is in the trace norm.
By \cite{object} this, in turn, means that the states $\ket a$ of the decohered system become objective as 
the information about the index $a$ has been redundantly copied and stored in the environment and is accessible without
disturbing the system through projective measurements on the supports of $\varrho_a^{mac}$.

\subsection{Model of Quantum Brownian Motion}

We begin with briefly describing the model \cite{Petruccione,Schlosshauer} and the results of \cite{qbm}. The central system $S$ is modeled as a harmonic oscillator of 
a mass $M$ and a frequency $\Omega$, linearly coupled to a bath of oscillators:
\begin{equation}\label{H}
\hat H=\frac{\hat P^2}{2M}+\frac{M\Omega^2\hat X^2}{2}+\sum_{k=1}^N\left(\frac{\hat p_k^2}{2m_k}
+\frac{m_k\omega_k^2\hat x_k^2}{2}\right)+\hat X\otimes\sum_{k=1}^N C_k\hat x_k,
\end{equation}
where we assume the units in which $\hbar=1$, $\hat x_k, \hat p_k$ describe the $k$-th environmental oscillator and $C_k$ are the coupling constants.
The central assumption is that the system is very massive compared to the environments: $M\gg m_k$ for every $k$.
The formation of the broadcast state is measured by two scalar functions: the well known decoherence factor, associated with
a loss of some of the environments and a generalized overlap \cite{Fuchs}, measuring distinguishability of macroscopic states of the observed 
environments. The whole system is initialized in a product state $\varrho_{0S}\otimes\bigotimes_k\varrho_{0k}$,
where about the central systems' initial state $\varrho_{0S}$ we only assume that it has coherences in the position $X$.
Apart from the last subsection, the environments are assumed  to be in a thermal state with the same temperature $T$.
The noise in the environment is a highly non-trivial factor, compared to pure initial states (as in e.g. \cite{qbm_Zurek}),
both conceptually, as the noise can prevent accumulation of information in the environment although decoherence has taken place,
and technically \cite{sfera,qbm}. We are interested in collective effects, where a single environment couples very weakly to the system.
Otherwise, it could "learn" enough about the system to decohere it---a situation trivial from a point of view of studying
information flow into different parts of the environment.
In QBM this means that the environmental oscillators are off-resonance: 
\begin{equation}\label{offres}
\omega_k\ll\Omega \ \ \text{or} \ \ \omega_k\gg\Omega\ \ \text{for all}\ k.
\end{equation}
Because a single environment acquires vanishingly small amount of information in the course of the interaction,
we have proposed \cite{sfera} to group the observed environments into fractions, scaling with the total number of the environments $N$, 
so called macrofractions. This is at the level of such coarse-grained environment where one can expect an appearance of a spectrum broadcast structure.
This  method is analogous to e.g. description of liquids: each point of a liquid (a macro-fraction here) is in reality composed of a suitable large number of 
microparticles (individual environments). Same technique is also used in a mathematical approach to von Neumann
measurements using, so called, macroscopic observables (see e.g. \cite{mo} and the references therein).

Another characteristic feature of our method \cite{qbm} is that instead of introducing the usual continuum approximation of the environments, characterized by a continuous
spectral density function \cite{Schlosshauer,Petruccione}, we work with a discrete bath of oscillators with random frequencies $\omega_k$ 
(this method is adapted from the spin systems \cite{Zurek_spins}).
Under such conditions, we have obtained analytical expressions for the decoherence  and distinguishability factors, 
denoted $\Gamma_{X,X'}(t)$ and  $B^{mac}_{X,X'}(t)$ respectively, between two different initial positions $X,X'$ (or in fact states of motion in a certain sense)
of the central system \cite{qbm}:
\begin{eqnarray}
&&\left| \Gamma_{X,X'}(t)\right|=\exp\left[-\frac{(X-X')^2}{2}\sum_{k\in(1-f)E}|\alpha_k(t)|^2\text{cth}\left(\frac{\beta\omega_k}{2}\right)\right],\label{GT}\\
&&B^{mac}_{X,X'}(t)=\exp\left[-\frac{(X-X')^2}{2}\sum_{k \in mac}|\alpha_k(t)|^2\text{th}\left(\frac{\beta\omega_k}{2}\right)\right],\label{BT}\\
&&|\alpha_k(t)|^2=\frac{C_k^2\omega_k}{2m_k(\omega_k^2-\Omega^2)^2}
\bigg[\left(\cos\omega_kt-\cos\Omega t\right)^2+\left(\sin\omega_kt-\frac{\Omega}{\omega_k}\sin\Omega t\right)^2\bigg],\label{aT}
\end{eqnarray}
where $(1-f)E$ is the unobserved (traced out) fraction of the environment, $mac$ denotes one of the observed macrofractions,
and $\beta=1/k_BT$ is the inverse temperature. Due to the random character of $\omega_k$'s these are complicated almost periodic functions of time.
Numerical analysis in \cite{qbm}, assuming independently, identically distributed (i.i.d.) $\omega_k$'s with a uniform distribution over a
spectrum interval far from the resonance, shows that if the macrofraction sizes are large enough for a given temperature, the broadcast structure is formed,
certified by vanishing of both of the above functions. 
These results are in accord with the earlier results of \cite{qbm_Zurek}, which suggested there is  indeed some objectivisation in the model, 
in the sense that we confirm the objectivisation but using a very 
different, more fundamental approach \cite{object}, based directly on quantum states and for a much wider and more realistic class of the environment 
initial states---thermal states. 
Below, using various simplifications, we further study formation of the structure (\ref{br2}) analytically as well
as the effect of squeezing in the initial thermal state.

\subsection{Full Quantum Measurement Limit And Exact Timescales}

We first consider a highly simplified model, with dominating interaction term and completely 
neglecting self-Hamiltonians of both the system and the environment:
\begin{equation}
\hat H\approx \hat X\otimes \sum_{k=1}^N C_k\hat x_k
\end{equation}
This is called Quantum Measurement Limit due to the fact that the  Hamiltonian represents an ideal von Neumann measurement
of the system observable $\hat X$ by the environments \cite{vN}. Such a simplified model will allow us to obtain the character and the timescales
of the broadcast structure formation.
The corresponding evolution is of the controlled unitary type with $X$ as the control parameter:
\begin{equation}
\hat U(t)=\int dX \ket X \bra X \otimes \bigotimes_{k=1}^N \hat D \left(-\frac{iC_kt}{\sqrt 2}X\right),
\end{equation}
where $\hat D(\alpha)=e^{\alpha a^\dagger-\alpha^*a}$ is the displacement operator and we have set the  oscillators masses and the
frequencies to unity. This leads to the following decoherence and distinguishability factors \cite{qbm}:
\begin{eqnarray}
&&\left| \Gamma_{X,X'}(t)\right|=\exp\left[-\frac{(X-X')^2}{2}t^2\text{cth}\frac{\beta}{2}\sum_{k\in(1-f)E} C_k^2\right],\\
&&B^{mac}_{X,X'}(t)=\exp\left[-\frac{(X-X')^2}{2}t^2\text{th}\frac{\beta}{2}\sum_{k\in mac} C_k^2\right].
\end{eqnarray}
One immediately sees that unlike in the full model, in the quantum measurement limit there is always a formation
of the spectrum broadcast structure, and hence objectivisation of the system's position $X$, described by the Gaussian decay in time
of $\left| \Gamma_{X,X'}(t)\right|$, $B^{mac}_{X,X'}(t)$.
Quite surprisingly, one does not even have to assume random coupling constants $C_k$ as e.g. in the spin systems \cite{Zurek_spins}.
If, however, one assumes random i.i.d. $C_k$'s with a finite average $\overline{C^2}<\infty$,  
more can be said as the sum $\sum_k C_k^2$ above can be approximated using  the law of large numbers for large macrofraction sizes:
\begin{eqnarray}
&&\left| \Gamma_{X,X'}(t)\right|\underset{N\to \infty}\approx\exp\left[-(1-f)N\frac{(X-X')^2}{2}\overline{C^2}t^2\text{cth}\frac{\beta}{2}\right]
=e^{-(1-f)N\left(\frac{t}{\tau_D}\right)^2},\label{Gg}\\
&&B^{mac}_{X,X'}(t)\underset{N\to \infty}\approx\exp\left[-mN\frac{(X-X')^2}{2}\overline{C^2}t^2\text{th}\frac{\beta}{2}\right]=
e^{-mN\left(\frac{t}{\tau_B}\right)^2},\label{Bg}
\end{eqnarray}
where $(1-f)N$ and $mN$, $f,m\in(0,1)$, are the sizes of the unobserved macrofraction $(1-f)E$ and (one of the) observed one $mac$ 
respectively. The characteristic timescales of decoherence and distinguishability are given by, respectively:
\begin{eqnarray}
&&\frac{1}{\tau_D}=\left|X-X'\right|\sqrt{\frac{\text{cth}(\beta/2)}{2}\overline{C^2}},\\
&&\frac{1}{\tau_B}=\left|X-X'\right|\sqrt{\frac{\text{th}(\beta/2)}{2}\overline{C^2}},
\end{eqnarray}
with the second being in general larger than the first for a given temperature $T$. This is the effect of noise---it obviously slows down 
the accumulation of information in environmental macrofractions (cf. \cite{sfera}).

\subsection{Partial Quantum Measurement Limit And Time Averages}

Inclusion of the self-Hamiltonians of the environment:
\begin{equation}
\hat H=\sum_{k=1}^N\left(\frac{\hat p_k^2}{2m_k}+\frac{m_k\omega_k^2\hat x_k^2}{2}\right)+\hat X\otimes\sum_{k=1}^N C_k\hat x_k,
\end{equation}
still allows for an exact solution, following e.g. the method of \cite{qbm}:
\begin{eqnarray}
&&\hat U(t)=\int dX \ket X \bra X \otimes \bigotimes_{k=1}^N \hat D\left(\alpha_k(t)X\right) e^{i\zeta_k(t) X^2},\\
&&\alpha_k(t)= -\frac{C_k}{\sqrt{2m_k\omega_k^3}}\left[e^{i\omega_kt}-1\right], \ 
\zeta_k(t)= \frac{C_k^2}{ m_k \omega_k^3} \left[\omega_kt -\sin(\omega_kt)\right].
\end{eqnarray}
For random $\omega_k$'s, the decoherence and distiguishability factors become
almost periodic functions of time:
\begin{eqnarray}
&&\left| \Gamma_{X,X'} \right| =\exp\left[\frac{(X-X')^2}{2}\sum_{k\in (1-f)E}\text{cth}\left(\frac{\beta\omega_k}{2}\right)
\frac{C_k^2\left(\cos \omega_k t-1\right)}{m_k \omega_k^3}\right] ,\label{G2}\\
&& B^{mac}_{X,X'}(t)= \exp\left[\frac{(X-X')^2}{2}\sum_{k\in mac}\text{th}\left(\frac{\beta\omega_k}{2}\right)
\frac{C_k^2\left(\cos \omega_k t-1\right)}{m_k \omega_k^3}\right]. \label{B2}
\end{eqnarray}
They are simpler than in the full model \cite{qbm}, as e.g. there is no frequency mixing between the environmental bands and the central system,
but still too complicated for an immediate analytical studies (which will be the subject of an independent work). 
We believe the law of large numbers will prove useful again here. What however can be evaluated are the time averages
$\left\langle |\Gamma_{X,X'}| \right\rangle = \lim_{\tau\to\infty}\frac{1}{\tau} \int^{\tau}_{0} dt |\Gamma_{X,X'}(t)|$, 
$\left\langle B^{mac}_{X,X'}\right\rangle = \lim_{\tau\to\infty}\frac{1}{\tau} \int^{\tau}_{0} dt B^{mac}_{X,X'}(t)$,
noting that the amplitudes of the cosines in (\ref{G2},\ref{B2}) are all non-negative. Using a theorem 
from \cite{Wi} one obtains (see Methods):
\begin{eqnarray}
&&\left\langle \left| \Gamma_{X,X'} \right| \right\rangle=\exp\left[-\frac{(X-X')^2}{2} \sum_{k\in (1-f)E}  
\frac{C_k^2\text{cth}(\beta\omega_k/2)}{m_k \omega_k^3}\right] \prod_{k\in (1-f)E} \text{I}_0\left[\frac{(X-X')^2C_k^2\text{cth}(\beta\omega_k/2)}{2m_k \omega_k^3}\right],\nonumber\\
&&\label{G3}\\
&&\left\langle B^{mac}_{X,X'}\right\rangle=
\exp\left[-\frac{(X-X')^2}{2} \sum_{k\in mac}  \frac{C_k^2\text{th}(\beta\omega_k/2)}{m_k \omega_k^3}\right] \prod_{k\in mac}  
\text{I}_0\left[ \frac{(X-X')^2C_k^2\text{th}(\beta\omega_k/2)}{2m_k \omega_k^3}\right],\label{B3}
\end{eqnarray}
where $\text I_0(z)=(1/\pi)\int_0^\pi d\theta e^{z\cos\theta}$ is the modified Bessel function of the first kind.
The rationale behind studying time averages is that since the functions themselves are non-negative,
vanishing of the time averages is a good measure of the functions being practically zero. 
We are again interested in the scaling of the averages with the macrofraction sizes, for simplicity
assuming them here to be equal: $\#(1-f)E=\#mac=mN$, $m\in(0,1)$. We also assume $C_k$'s to be only 
mass-dependent (and not random) \cite{qbm}, $C_k=\sqrt{\frac{M m_k \tilde\gamma_0 }{\pi}}$, with $M$ the mass of the central system and 
$\tilde\gamma_0$ some constant. The scaling can be evaluated in 
the large separation limit:
\begin{equation}
\frac{\sqrt{M\tilde\gamma_0}|X-X'|}{\omega_k^{3/2}}\gg 1 \ \  \text{for every} \ k, \label{L} 
\end{equation}
which allows one to use the asymptotic expansion $\text I_0(x)\approx e^x/\sqrt{2\pi x}$.
Assuming further low temperature limit, $\beta\to\infty$ (from (\ref{B2}) $B^{mac}_{X,X'}$ rises with $T$), one obtains:
\begin{eqnarray}
\left\langle \left| \Gamma_{X,X'} \right| \right\rangle\approx \prod_{k=1}^{mN} \frac{\omega_k^{3/2}}{\sqrt{M\tilde\gamma_0}|X-X'|},
\end{eqnarray}
and the same for $\left\langle B^{mac}_{X,X'}\right\rangle$, as $\text{th}(\beta\omega_k/2)\approx 1 \approx \text{cth}(\beta\omega_k/2)$ in (\ref{G3},\ref{B3}).
We perform further averaging over the random frequencies $\omega_k$, assuming \cite{qbm}
they are i.i.d. with a uniform distribution over a spectrum interval $\Delta$, centered at some $\bar\omega\gg\Delta$. This gives:
\begin{equation}
\overline{\left\langle \left| \Gamma_{X,X'} \right| \right\rangle}=\int_{\bar\omega-\frac{\Delta}{2}}^{\bar\omega+\frac{\Delta}{2}}
\prod_j\frac{d\omega_j}{\Delta}\left\langle \left| \Gamma_{X,X'} \right| \right\rangle\approx 
e^{-mN\left(\log\frac{\sqrt{M\tilde\gamma_0}|X-X'|}{\bar\omega^{3/2}}\right)}
\end{equation}
and the same for the distinguishability factor, due to the assumed low temperature limit: 
\begin{equation}
\overline{\left\langle B^{mac}_{X,X'}\right\rangle}=\int_{\bar\omega-\frac{\Delta}{2}}^{\bar\omega+\frac{\Delta}{2}}
\prod_j\frac{d\omega_j}{\Delta}\left\langle \left|  B^{mac}_{X,X'} \right| \right\rangle\approx 
e^{-mN\left(\log\frac{\sqrt{M\tilde\gamma_0}|X-X'|}{\bar\omega^{3/2}}\right)}.
\end{equation}
This is the desired scaling.
One sees that whenever (\ref{L}) holds, for low temperatures, both averages exponentially decay 
in the thermodynamic limit $N\to\infty$, indicating formation of the spectrum broadcast state.

\subsection{Squeezed Thermal States in The Full Model}

\begin{figure}[h!]
\centering

\subfloat[ 10 oscillators]{\includegraphics[width=0.45\textwidth]{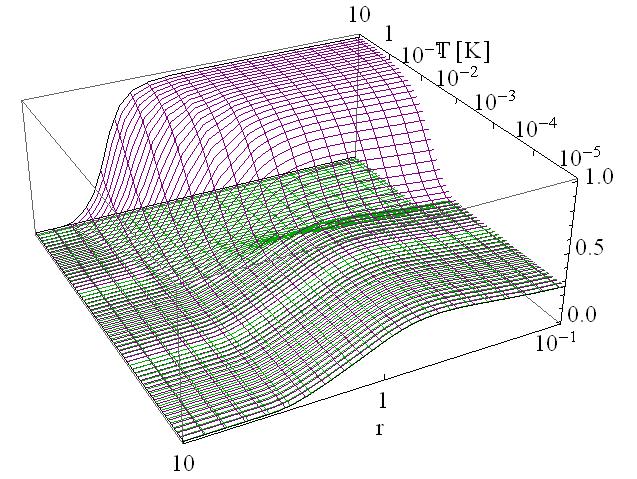}\label{10}}\\
\subfloat[ 30 oscillators]{\includegraphics[width=0.45\textwidth]{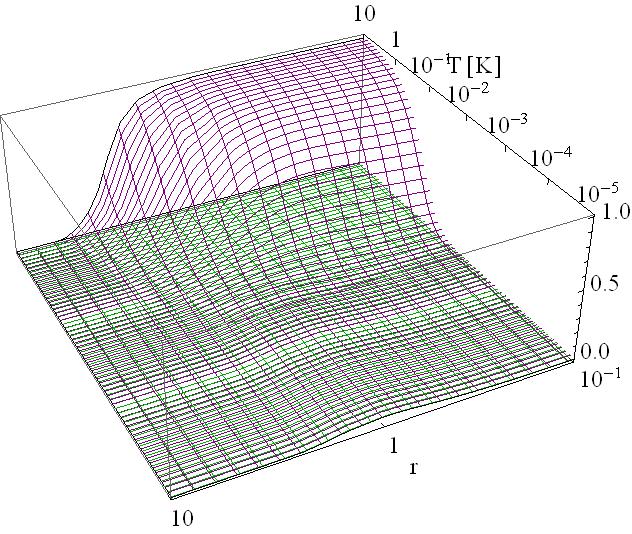}\label{30}}

\caption{ Time averaged decoherence $\langle|\Gamma_{X,X'}|\rangle$ (green; lower surfaces) and 
distinguishability $\langle B^{mac}_{X,X'}\rangle$ (magenta; upper surfaces) factors 
as functions of the temperature $T$ and squeezing $r$ (both on the logarithmic scale)
of the environment. The set of random frequencies $\omega_k$ was generated once per plot.
The traced over and observed macrofractions, $(1-f)E$ and $mac$, are assumed to be of the same size. 
Both plots simultaneously approaching zero  
indicates formation of the spectrum broadcast structure for a given $(r,T)$.
}\label{plot}
\end{figure}

We now switch to the full model (\ref{H}) and analyze the effect of squeezing in the initial thermal state of the environment, assuming:
\begin{equation}
\varrho_{0k}=\hat S(r)\varrho_{kT}\hat S(r)^\dagger,
\end{equation}
where $\hat S(r)=e^{\tfrac{r}{2}(\hat a ^2-\hat a^{\dagger 2})}$ is the squeezing operator and $\varrho_{kT}$ are the thermal oscillator states. 
Simple calculation shows that it is enough to substitute in (\ref{GT},\ref{BT}) $|\alpha_k(t)|^2$ for:
\begin{eqnarray}
&&|\tilde\alpha_k(t)|^2=\text{ch} (2r) \left[\left|\alpha_k(t)\right|^2  - \text{th} (2r) \text{Re} \alpha_k^2 (t)\right],\label{ar}\\
&&\text{Re} a_k(t)^2 = \frac{C_k^2}{4 m_k \omega_k}\bigg\{\frac{1}{(\omega_k+\Omega)^2}\left\{\cos\left[2(\omega_k+\Omega)t\right] - 2 \cos\left[(\omega_k+\Omega)t \right] \right\} \nonumber\\
&&+ \frac{1}{(\omega_k-\Omega)^2}\left\{\cos\left[2(\omega_k-\Omega)t \right] - 2 \cos\left[(\omega_k-\Omega)t\right] \right\} \nonumber\\ 
&& + \frac{1}{\omega_k^2-\Omega^2}\left\{\cos2 (\omega_k t) - \cos\left[\left(\omega_k-\Omega \right)t\right]  - \cos\left[\left(\omega_k+\Omega \right)t\right]  \right\} + \frac{3 \omega_k^2+ \Omega^2}{(\omega_k^2-\Omega^2)^2} \bigg\}. 
\label{ar2}
\end{eqnarray}
From (\ref{ar}) one sees that for large squeezing, $|\tilde\alpha_k(t)|^2$ grows exponentially as $e^{2r}$, 
enabling a formation of the broadcast structure as the most problematic term in (\ref{BT}) decay as $\text{th}(\beta\omega_k/2)\approx \beta\omega_k/2$ for 
high temperatures. 
This reflects the fact that large squeezing decreases the noise, increasing the information capacity of the environment.
The plots in Fig. \ref{plot} confirm this. 
From Fig. \ref{10} it can be seen that for a moderate macrofractions of 10 oscillators 
only with a help of a large squeezing  $\log r >0$ the negative effect of the temperature
can be overcome, and both functions are damped  indicating formation of the spectrum broadcast structure.

For 30 oscillators, Fig. \ref{30}, the effect of random phases in (\ref{GT},\ref{BT}) is much stronger, the functions are much more damped for the chosen parameters
range, and hence the plateau of a broadcast structure formation is much larger. Quite surprisingly, there is a bump
around $\log r=0$ where the formation is suppressed, most probably due to a constructive interference
in (\ref{ar},\ref{ar2}).

The parameters for the numerics are as follows \cite{qbm}: For the central system we assume $M=10^{-5}$ kg, $\Omega=3 \times 10^{8}$ s$^{-1}$;
the frequencies $\omega_k$ are i.i.d. and uniformly distributed in the interval $3\dots 6 \times 10^{9}$ s$^{-1}$ to satisfy
the off-resonance condition (\ref{offres}); we set $|X-X'| = 10^{-9}$m.
We assume here that the coupling constants $C_k$ depend only on the masses,
$C_k\equiv 2\sqrt{\frac{M m_k \tilde\gamma_0 }{\pi}}$, where $\tilde\gamma_0 =0.33\times 10^{18}$  s$^{-2}$ is a constant.
Thus, we do not assume the dependence on the $\omega_k$'s, as it is the case e.g. for the Ohmic spectral density \cite{qbm_Zurek}.
 
The averaging time was set at fairy large $1s$.
Finally, as in the previous section, we assume a symmetric situation where the size of the traced macro-fraction $(1-f)E$ in (\ref{GT}) is the same as the size of 
the observed one $mac$ in (\ref{BT}).

\section{Discussion}

Our results can be grouped into three groups. First, neglecting the self-Hamiltonians
of the system and the environments completely we have shown that there is always a formation of the spectrum broadcast structure
for thermal environments, irrespectively of how high the temperature is, if one waits long enough and/or takes large enough macrofractions.
The derived Gaussian character of this formation together with characteristic timescales has not been obtained before. An application of the law of large numbers allowed us to obtain the exponential scaling
of the decay with the environment size. Although very basic here, we believe this method will make analytical
studies of the full model possible.

Second, in more complicated model including self-Hamiltonians of the environments we have shown using 
a form of ergodicity for almost periodic functions \cite{Wi} an exponential decay
of infinite-time averages of both decoherence and distinguishability factors in the low temperature limit and for large separations
of the systems' initial positions $X,X'$. Again, this is a novel result, compared to the earlier studies \cite{qbm_Zurek,qbm}.
The large separation assumption opens here an interesting possibility of a space coarse-grained spectrum broadcast structure,
where such a structure appears for large distances only but for small not. This would lead to a sort of a {\em macroscopic objectivity},
emerging only at large scales. This concept will be a subject of a further research.

Finally, we have numerically studied the effect of squeezing in the thermal environment in the full model. 
To our knowledge such studies have not been performed before in the context of objectivisation 
(in \cite{qbm_Zurek} the system was initialized in a squeezed state, but the environment was initially in the ground state).
Squeezing obviously increases the informational capacity of the environment, while the temperature destroys it.
This leads to an interplay between squeezing and heating of the environment and our results show that
in general large enough squeezing will dominate the noise due to the temperature. We had to rely on the
numerics as the appearing almost periodic functions are quite complicated. However, as mentioned above,
we hope the law of large numbers together with possible additional simplifications, e.g. the low temperature limit,
 will open a way for obtaining analytical estimates on the timescale of the objectivisation.

\section{Methods}

\subsection{Time averages of almost periodic functions with positive coefficients}

We will show the passage from (\ref{G2},\ref{B2}) to (\ref{G3},\ref{B3}).
We evaluate the time average:
\begin{eqnarray}
&&\lim_{\tau \to \infty}\frac{1}{\tau} \int_{0}^{\tau} d t  \exp\left[\frac{(X-X')^2}{2}\sum_{k\in (1-f)E}\frac{C_k^2\cos \omega_k t }{m_k \omega_k^3}\text{cth}\left(\frac{\beta\omega_k}{2}\right)\right].
\end{eqnarray}
The crucial step is to use one of the results  from \cite{Wi}, which states a form of ergodicity for such functions. Namely,  due to the positive amplitudes of the trigonometric functions, 
the time average can be substituted with the ensemble average (average over the angles in this case). This gives:
\begin{eqnarray}
&& \lim_{\tau \to \infty}\frac{1}{\tau} \int_{0}^{\tau} d t \prod_{k\in (1-f)E}  \exp\left[\frac{(X-X')^2}{2}\frac{C_k^2\cos \omega_k t }{m_k \omega_k^3}\text{cth}\left(\frac{\beta\omega_k}{2}\right)\right] \\
&& =\prod_{k\in (1-f)E} \frac{1}{2 \pi} \int_{0}^{2 \pi} d \theta_k  \exp\left[\frac{(X-X')^2}{2}\frac{C_k^2 }{m_k \omega_k^3}\text{cth}\left(\frac{\beta\omega_k}{2}\right) \cos \theta_k \right]\\
&& =\prod_{k\in (1-f)E} \text{I}_0\left[\frac{(X-X')^2}{2}\frac{C_k^2}{m_k \omega_k^3}\text{cth}\left(\frac{\beta\omega_k}{2}\right) \right],
\end{eqnarray}
where the last step is the standard Bessel integral, giving the modified Bessel function of the first kind: $\text I_0(z)=(1/\pi)\int_0^\pi d\theta e^{z\cos\theta}$.

\acknowledgments{Acknowledgments}

We would like to thank Ryszard Horodecki, Pawe\l \, Horodecki, and especially Jan Wehr
for insightful discussions.
This work is supported by the ERC Advanced Grant QOLAPS and  
National Science Centre project Maestro DEC-2011/02/A/ST2/00305. 

%%%%%%%%%%%%%%%%%%%%%%%%%%%%%%%%%%%%%%%%%%

\section{Author Contributions}

All authors contributed to all aspects of this work.

%%%%%%%%%%%%%%%%%%%%%%%%%%%%%%%%%%%%%%%%%%

\section{Conflicts of Interest}

The authors declare no conflict of interest. 

%=================================================================
% References: Variant A
%=================================================================
% Back Matter (References and Notes)
%----------------------------------------------------------
% Style and layout of the references
\bibliographystyle{mdpi}
\makeatletter
\renewcommand\@biblabel[1]{#1. }
\makeatother

%=================================================================
% References:  Variant B
%=================================================================
% Use the following option to include external BibTeX files:
%\bibliography{lite}
%\bibliographystyle{mdpi}

%%%%%%%%%%%%%%%%%%%%%%%%%%%%%%%%%%%%%%%%%%

%\abbreviations{Abbreviations/Nomenclature}
%
%Main text.

%%%%%%%%%%%%%%%%%%%%%%%%%%%%%%%%%%%%%%%%%%

%\appendix
%\section{Appendix Title}
%
%Main text.

\end{document}